\documentclass[aps,twocolumn,groupedaddress,amssymb]{revtex4}
\usepackage{latexsym}
\usepackage{amsbsy}
\usepackage{amsmath}
\usepackage{graphicx}

\usepackage{color}

\usepackage{color}
\usepackage{hyperref}
\hypersetup{
    colorlinks=true,
    linkcolor=black,
    urlcolor=blue,
    citecolor=blue
}

\begin{document}
\title{Anomalous Hall effect in highly c-plane oriented Mn$_{3}$Ge/Si(100) thin films grown by pulsed laser deposition}

\author{
    Indraneel Sinha\textsuperscript{1}, 
    Purba Dutta\textsuperscript{2},
    Nazma Firdosh\textsuperscript{1}
    Shreyashi Sinha\textsuperscript{1}
    Nirmal Ganguli\textsuperscript{2}, 
    Sujit Manna\textsuperscript{1}
}
\email{smanna@physics.iitd.ac.in}

\affiliation{\textsuperscript{1}Department of Physics, {Indian Institute of Technology Delhi}, Hauz Khas, New Delhi 110016, India}
\affiliation{\textsuperscript{2}Department of Physics, {Indian Institute of Science Education and Research Bhopal}, Bhauri, Bhopal 462066, India}

\floatsep 0.03 in
\textfloatsep 0.03in
\intextsep 0.03 in
\dbltextfloatsep 0.03 in
\dblfloatsep 0.03 in
\abovecaptionskip 0.03 in
\belowcaptionskip 0.03 in


\begin{abstract}

 Antiferromagnetic Mn$_{3}$Ge with a non-collinear Kagome structures present exciting prospects for exploring Berry curvature driven anomalous Hall effects (AHE). 
 Despite substantial progress in bulk systems, the synthesis of crystalline thin films directly on silicon with a hexagonal phase presents a particular challenge unless a buffer layer is employed.
 In this study, we report the synthesis of single phase c-plane oriented  hexagonal Mn$_{3}$Ge(0001) films on Si(100) using pulsed laser deposition. Under suitable growth conditions, we obtain layer-by-layer films with atomically flat surfaces and interfaces. High-resolution scanning tunneling microscopy study reveals the detail surface atomic structures, where the surface Mn atoms spontaneously arrange into a Kagome lattice. Tunneling spectroscopy (dI/dV) measurement on the atomically resolved Kagome surface show a minima in local density of states near the Fermi level, likely originated from the Weyl crossings near K points. Despite the nearly vanishing magnetization, magnetotransport measurements in 30 nm $Mn_{3}$Ge(0001) films show anomalous Hall resistivity up to 0.41 ($\mu\Omega\cdot\text{cm}$) at 2 K. 
Our \textit{ab initio} calculations shed further light on the existence of topological features and the band structures in Mn$_{3+x}$Ge$_{1-x}$ with increasing Mn concentration $x$. The anomalous Hall response at room temperature in crystalline Mn$_{3}$Ge films on Si(100) offer promising potential for the development of antiferromagnetic spintronics.

\end{abstract}
\maketitle
\section{Introduction}
Non-collinear antiferromagnets (AFM) with chiral-spin ordering in $Mn_3X$ (where X = Ga, Ge, Sn) family materials are attracting significant research focus, with the promise of hosting a large anomalous Hall effect (AHE) at room temperature, despite their vanishingly small net magnetization \cite{nayak2016large,yang2017topological,chen2021anomalous,vsmejkal2022anomalous,kiyohara2016giant}. This has been suggested as a pathway to a number of novel quatum phenomena, including the topological Hall effect \cite{kimata2019magnetic}, the anomalous Nernst effect \cite{hong2020large}, and large magneto-optic Kerr rotation 
 \cite{wu2020magneto}. Kagome magnets also providing a unique platform for gaining fundamental insights into many-body electronic structure such as magnetic Weyl fermions, flat band and 
 high-order Van Hove singularities \cite{bernevig2022progress,classen2024high}. These quantum states arise from unusual geometry in the Kagome lattice and quantum interactions that lead to novel magnetism. 
 
 Both $Mn_3Ge$ and $Mn_3Sn$ share the hexagonal space group ($P6_{3}/mmc$) where Mn atoms arrange in a triangular geometry in the Kagome planes \cite{yang2017topological}. The novel non-collinear magnetic structure occurs due the geometrical frustration and the Dzyaloshinskii–Moriya interaction between the Mn spins at the corners of the triangles \cite{zhang2017strong,soh2020ground}.  Multiple Weyl points appear in reciprocal space due to the presence of mirror symmetry and time-reversal symmetry. These symmetries prevent the Weyl points with opposite chirality from annihilating each other, ensuring a non-vanishing Berry flux. A large AHE is observed in bulk $Mn_3Ge$ and $Mn_3Sn$ under an external magnetic field, originating from the non-vanishing integral of Berry curvature in momentum space \cite{nayak2016large,li2023field,yang2017topological,chen2021anomalous,kiyohara2016giant}. These topological non-trivial effects are significantly suppressed in thin films due to structural and compositional anomaly. The delicate relationship between the chiral magnetic ground state and chemical composition makes it challenging to observe AHE in thin films \cite{ikeda2020fabrication}. Till date, most experimental attempts to synthesize both epitaxial and polycrystalline films are focused on $Mn_3Sn$ \cite{gao2022epitaxial,khadka2020kondo,liu2023anomalous,taylor2020anomalous,ikeda2018anomalous,zhao2021magnetic,xie2022magnetization}. $Mn_3Ge$  emerges as a more favorable candidate than $Mn_3Sn$, owing to its comparatively higher anomalous Hall conductivity and the presence of higher number of Weyl points \cite{chen2021anomalous,kubler2018weyl}. Additionally, $Mn_3Ge$ retains its chiral spin order down to low temperature, unlike $Mn_3Sn$ which it transforms to a spin glass at 50 K \cite{sung2018magnetic}. 
In comparison to $Mn_3Sn$, Hexagonal ($DO_{19}$) and tetragonal ($DO_{22}$) phases of  Mn$_{3}$Ge are interchangeable through a structural deformation along the (111) direction \cite{manna2018heusler,kirste2023phase,zhang2013first}. The thermodynamics behind the stability of tetragonal and hexagonal phases in thin films is complex and influenced by a broad range of factors like  strain, film composition, and growth temperature \cite{berche2014thermodynamic}. Therefore, the synthesis of high-quality Mn$_{3}$Ge thin films on insulating substrates remains a challenge due to factors like wetting and diffusivity. 

\begin{figure*}
    \centering
    \includegraphics[width=1.0\textwidth]{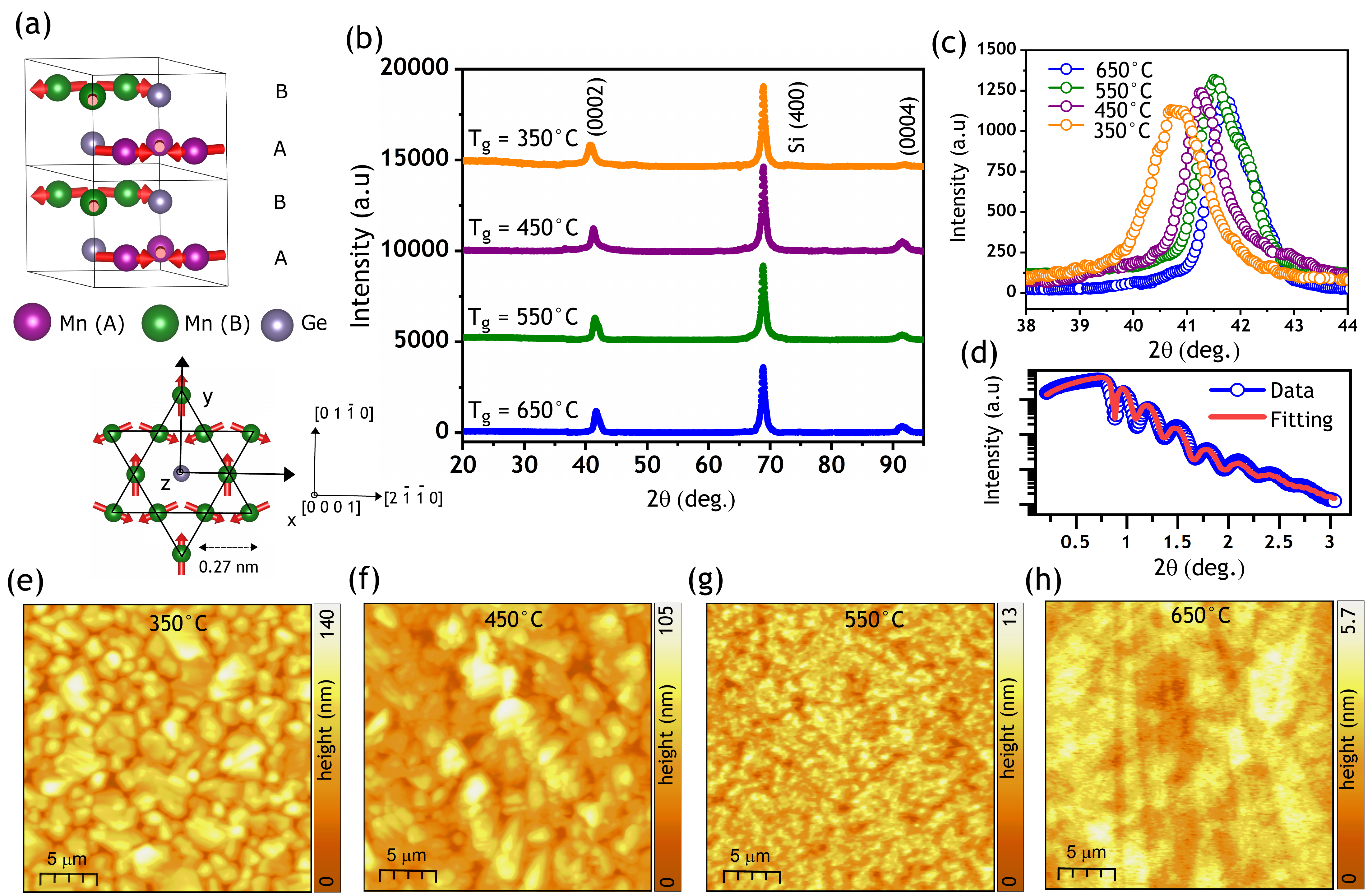} 
    \caption{Structural characterization of $Mn_{3}Ge$ films grown on Si(100). (a) Crystal structure of hexagonal Mn$_{3}$Ge phase showing two unit cells along the c axis (top) and projection of the basal plane (bottom). (b) x-ray diffraction patterns of Mn$_{3}$Ge thin films deposited at different temperatures show preferential growth along the (0001) axis on Si(100). (c) XRD pattern showing the detail of (0002) peak position of Mn$_{3}$Ge thin films grown at different temperatures. The shift in (0002) originates from the level of  Mn doping in $Mn_{3+x}Ge_{1-x}$. (d) The measured (open circle) and fitted (red line) x-ray reflectivity data of 30nm thick Mn$_{3}$Ge films grown at $650^{\circ}$ C. (e-h) Atomic force microscopy images (\(25 \times 25 \, \text{$\mu$m}^2\)) of  Mn$_{3}$Ge thin films grown at the four different temperatures ($T_{g}$ =) $350 ^{\circ}$ C, $450 ^{\circ}$ C, $550 ^{\circ}$ and C $650^{\circ}$ C respectively.}
    \label{fig:break1}
\end{figure*}

Initial attempts to induce exchange bias using polycrystalline $Mn_3Ge$ films showed negligible net  magnetization  {\cite{olayiwola2023room, ogasawara2019structural}, but the topological features of these films are not known. Wang et al. reported  the synthesis of both tetragonal ($DO_{22}$) and  hexagonal ($DO_{19}$) on  $SrTiO_{3}$ (001) substrate through post-annealing at different temperatures \cite{wang2021manipulation}. In a separate study conducted by  Wang et al., films grown at $400 ^{\circ}$ C had mixed tetragonal and hexagonal phases \cite{wang2022investigating}. Hong et al. utilized a three step synthesis process using Molecular Beam epitaxy (MBE) to grow epitaxial and continious  $Mn_3Ge$ ($DO_{19}$) films on $LaAlO_{3}$(111) substrate \cite{hong2022synthesis}. 
Moreover, studies show that the stability of $Mn_3Ge$  films directly on silicon are significantly limited without metallic buffer layers \cite{hong2020large,takeuchi2024magnetic}.  

  We show the details synthesis of highly order and perfectly crystalline Mn$_{3}$Ge(0001) films on Si(100) without any metallic buffer layers in between the films and substrate. Higher flux rates and controlled ablation from a single source Mn$_{3}$Ge target ensure a stoichiometric and layer-by-layer growth with well define interface. By tuning the growth parameters, we present a detail growth mechanism to engineer the composition and desire hexagonal phase over tetragonal \cite{bang2019structural} phase in Mn$_{3}$Ge films. 
In this study, we report a detail growth kinetics of single hexagonal phase using Scanning Tunneling Microscopy (STM). Additionally, a gap exists in the existing literature that focuses on STM studies on Kagome magnets, especially in the 2D-regime \cite{xing2020localized,li2022spin}. However, most novel phenomena like flat bands \cite{lin2018flatbands}, massive Dirac fermions \cite{ye2018massive}, and Van Hove singularities \cite{jiao2019signatures,zhang2023visualizing} emerge at the Kagome layer. Using scanning tunneling spectroscopy, we also gain insights into the spatial variation of electronic states on the Kagome plane.
We also theoretically investigate the effect of Mn doping by substituting Ge in Mn$_3$Ge with varying doping concentration. Our calculations considering spin-orbit interaction reveals substantial and collective change in electronic, magnetic and topological properties with increasing doping concentration. This study effectively suggests a transition from an antiferromagnetic conducting topological phase to a semiconducting phase without the Weyl nodes by introducing local inhomogeneties.

\section{Experimental methods}

Mn$_{3}$Ge thin films were grown on single crystalline Si(100) substrate by pulsed laser deposition at base pressure below $1\times 10^{-6} $ mbar. All these films were synthesized without using any metallic buffer layer from a single-source Mn rich  target with a pulsed KrF excimer laser source of wavelength 248 nm \cite{sinha2024magnetic}. The target-substrate distance was maintained at a constant 4.5 cm during deposition, with the laser fluence fixed at 3.5 $J cm^{-2}$ and a pulse frequency of 5 Hz set throughout the process. Si(100) substrates were prepared by wet-chemical etching followed by thorough degassing and a repeated cycle of in-situ annealing. A series of (0001) oriented crystalline films were prepared by varying the growth temperature between $350^{\circ}$ C to $650^{\circ}$ C, followed by one hour of post-annealing at the deposition temperature.

Structural characterization was performed using X-ray diffraction (XRD) with a PANalytical X'Pert diffractometer with Cu-K$\alpha$ source ($\lambda = 1.5418 \, \text{\AA}$). The thickness of the films were calibrated using X-ray reflectivity (XRR) measurements. A growth rate of 0.04 \text{\AA} per laser shot was verified, to establish a standard between thickness and growth time. The surface morphology was analyzed using atomic force microscopy (AFM) (Oxford Instruments Asylum Research, MFP-3D system). The elemental composition of  Mn$_{3}$Ge films were determined using energy-dispersive X-ray analysis (Hitachi Table Top TM3000). Scanning tunneling microscopy and spectroscopy (STM/STS) measurements were performed at room temperature using Quaza STM with a home-built integrated active vibration isolation system. Both commercial PtIr tip and annealed W tips followed by field emission was used as probe. Differential conductance ($dI/dV$) spectra were measured in STS mode using a lock-in amplifier by adding an AC (1.7 kHz) modulation voltage $V_{\text{rms}} = 20 \text{mV}$ to the bias voltage. All the topography were obtained in constant current mode \cite{sinha2024magnetic}. The magnetization of the Mn$_{3}$Ge films were measured using Quantum Design magnetic property measurement system (MPMS-3). Temperature and field-dependent magnetotransport measurements are performed in a standard four-probe geometry using a cryogen-free Physical Property Measurement System (PPMS).
The theoretical band structures were calculated using density functional theory (DFT) within the framework of the {\scshape vasp} code \cite{kresse1999ultrasoft}, employing a plane wave basis set with 500~eV cutoff energy in conjunction with the projector augmented wave (PAW) method \cite{blochl1994projector}. The exchange-correlation functional was described through the generalized gradient approximation (GGA) due to Perdew-Burke-Ernzerhof (PBE) \cite{perdew1996generalized}}. A $\Gamma$-centered $\vec{k}$ mesh of $3 \times 3 \times 11$ was employed for the integration over the Brillouin zone, using tetrahedron method \cite{blochl1994improved}. To account for the strong onsite Coulomb repulsion stemming from the localized Mn-3d electrons, we adopted DFT+$U$ approach\cite{dudarev1998electron}, with an effective Hubbard parameter ($U_{\text{eff}}$) of $3$~eV.

\begin{figure*}
    \centering
 \includegraphics[width=1.0\textwidth]{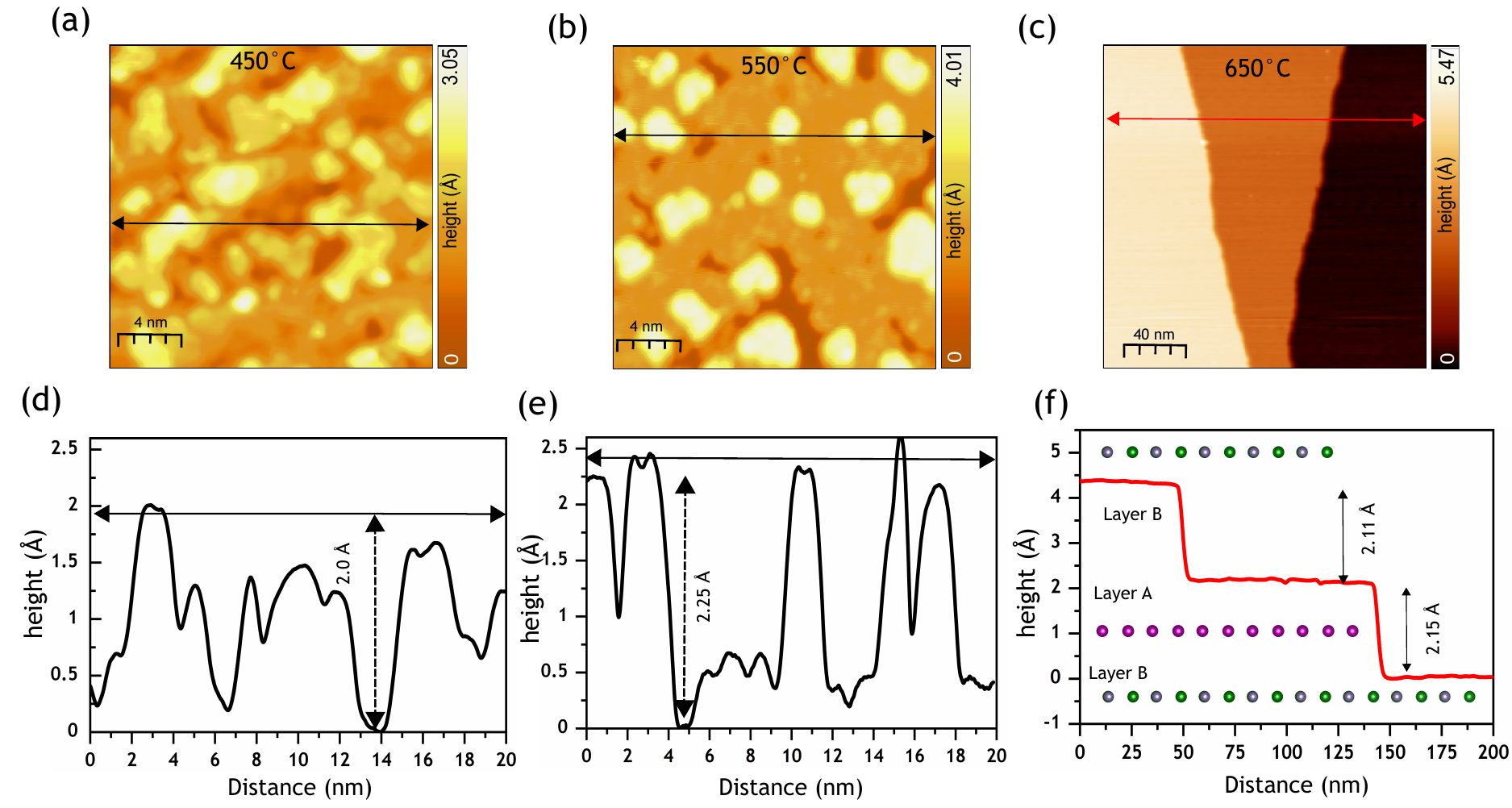} 
    \caption{ Constant current STM images of 30 nm thick Mn$_{3}$Ge films deposited at (a) $450^{\circ}$C, (b) $550^{\circ}$C, and (c) $650^{\circ}$C, respectively. (a) STM topographic image (\(20 \times 20 \, \text{nm}^2\)) taken on the surface of Mn$_{3}$Ge (0001) films grown at $450^{\circ}$C. The islands grow to a size of 5–10 nm, with multiple terraces resulting from a higher nucleation density. A bias voltage $V_{Bias}$ = 1.2 V and a tunneling current $I_{t}$ = 450 pA were used during the imaging. (b) Topographic image  \(20 \times 20 \, nm^2\) of Mn$_{3}$Ge (0001)/Si(100) grown at $550^{\circ}$C. In this temperature regime, film morphology transitions into a layer-by-layer growth mode with distinct terraces. Tunneling condition kept at $V_{Bias}$ = 1 V and $I_{t}$ = 210 pA. (c) Large area STM topography (\(200 \times 200 \, \text{nm}^2\)) obtain( $V_{Bias}$ = 0.5 V, $I_{t}$ = 100 pA) on the films deposited at $650^{\circ}$C. At this growth condition, the island merges to form atomically flat terraces with (0001) termination. The flat regions in the image correspond to the two possible terminations of the c-plane ( Layer A and Layer B).  (d) The line profile along the black line in (a) shows the height distribution of the facets. The height of the islands is determined from the individual peak profile. 
(e) The height of the peaks in the line profile (b) is approximately 2~\AA, close to the height difference between A and B layers along the c-axis.  
(f) The line profile taken along the red line in (c) shows three distinct terraces with an average width of 100 nm. The evolution of STM topography indicates that higher surface mobility at high deposition temperatures enhances terrace widths and favors a layer-by-layer growth mode. 
 }
    \label{fig:break2}
\end{figure*}
\section{Results and Discussions}
\subsection{Mode of growth and structural characterization}
 Despite advances in synthesizing and studying magnetotransport in bulk Mn$_{3}$Ge, the relationship between structural parameters, magnetic ordering, and transport properties in thin films remains unresolved. Previous reports show discrepancies regarding the influence of growth conditions for ($DO_{19}$) hexagonal phase and ($DO_{22}$) tetragonal phases \cite{kren1970neutron,ohoyama1961new}. However, thin films show a structural transition from hexagonal to tetragonal  with an increase in annealing temperature \cite{wang2021manipulation}. In our experiment, we vary the growth temperature between $350^{\circ}$ C to $650^{\circ}$ C to suppress any contribution from ($DO_{22}$) phase. 
 The crystal structure with varying growth temperatures are investigated using X-ray diffraction (XRD).
\hyperref[fig:break1]{\textcolor{blue}{Fig.~\ref*{fig:break1}(b)}} confirms the growth of single phase ($DO_{19}$) structure with strong characteristic peaks from (0002) and (0004) planes. As the growth temperature is changed from $350^{\circ}$ C to $650^{\circ}$ C, the primary (0002) peak shifted gradually to higher angles. This slight deviation is attributed to the varying composition in the films. During growth, it is highly probable of Mn atoms to occupy Ge sites, as  $Mn_{3+x}Ge_{1-x}$ is stable over a broad spectrum of x. The c-axis lattice constant determined from the (0002) peak positions in \hyperref[fig:break1]{\textcolor{blue}{Fig.~\ref*{fig:break1}(c)}} changes between  4.42 \AA ($350^{\circ}$ C) to 4.32 \AA ($650^{\circ}$ C) respectively. This suggests that the films are fully relaxed at $650^{\circ}$ C, approaching its bulk lattice constant \cite{qian2014exchange}. 
The compositional change with growth temperature is studied using EDX as shown in the supplementary section (Fig. S1). Previous reports show that the stability window for epitaxial to the polycrystalline regime is also governed by Mn-Ge composition \cite{yoon2021correlation}. In our case, a highly c-axis oriented growth is observed between $350^{\circ}$ C to $650^{\circ}$ C. 
The XRR profile of a film deposited at $650^{\circ}$ C is displayed in \hyperref[fig:break1]{\textcolor{blue}{Fig.~\ref*{fig:break1}(d)}}. The simulated model fits well with the intensity modulations for 30 $\pm$ 0.5 nm thick films with roughness less than 1 nm.  From the electrical viewpoint, topological aspects like Hall conductivity or mobility are strongly influenced by morphological properties like surface roughness or grain distribution. 
In our observation, films grown between $350^{\circ}$ C - $450^{\circ}$ C are highly insulating or in the semiconducting regime. $Mn_3{Ge}$ films start showing their typical metallic behavior at a  growth temperature of $550^{\circ}$ C. The longitudinal resistivity ($\rho_{xx}$) at 300 K exhibit an abrupt drop from 890 $\mu \Omega$cm to 187 $\mu \Omega$cm for $550^{\circ}$ C and $650^{\circ}$ C respectively.  AFM topographs shown in \hyperref[fig:break1]{\textcolor{blue}{Fig.~\ref*{fig:break1}(e)}} and \hyperref[fig:break1]{\textcolor{blue}{Fig.~\ref*{fig:break1}(f)}} were characterized on films grown at $350^{\circ}$ C and  $450^{\circ}$ C respectively. At this temperature range, films demonstrate isolated growth with discontinuity in between the islands. The semiconducting nature of these samples could originate from the diffused scattering at the grain boundaries. AFM topograph acquired on films deposited at $550^{\circ}$ C indicate the beginning of a continuous surface. Based on \hyperref[fig:break1]{\textcolor{blue}{Fig.~\ref*{fig:break1}(g)}} and \hyperref[fig:break1]{\textcolor{blue}{Fig.~\ref*{fig:break1}(h)}}, it is clearly identifiable that depositions around $550^{\circ}$ C - $650^{\circ}$ C show a shift towards a layer-by-layer growth with atomically flat surfaces. Analyzing these AFM images, it is visible that high-growth temperatures can assist the nucleation of grain boundaries due to enhanced surface mobility. 

\begin{figure*}
    \centering
    \includegraphics[width=1.0\textwidth]{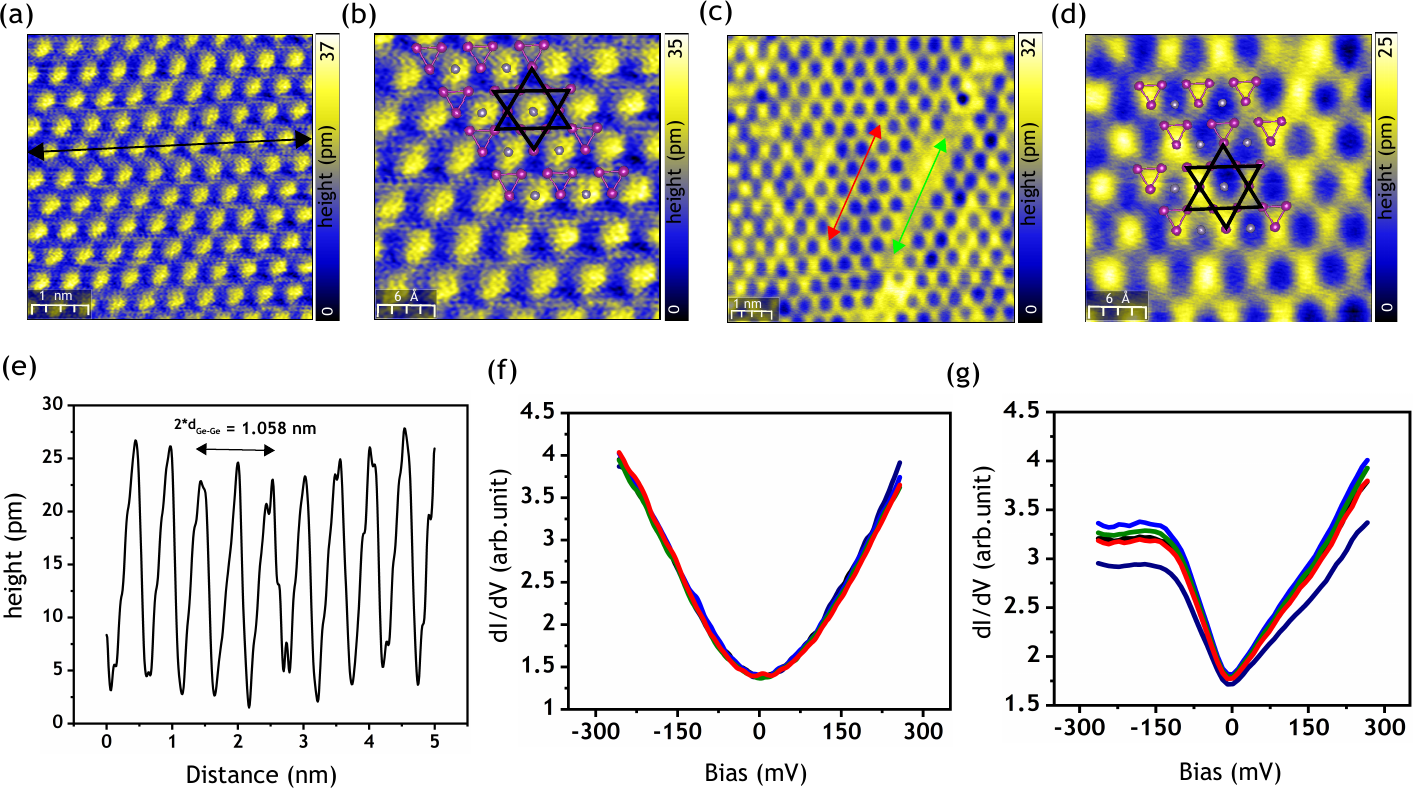} 
    \caption{ Atomically resolved Topographic image of 30nm thick Mn$_{3}$Ge thin films deposited at $650^{\circ}$ C. (a) high-resolution STM image \(5 \times 5 \, \text{nm}^2\) of the Kagome layers exhibiting the hexagonal closed packed geometry of Ge atoms($V_{Bias}$ = -125 mV, $I_{t}$ = 850 pA). (b) Zoomed-in image of (a) \(3 \times 3 \, \text{nm}^2\) with ball-and-stick models of the  Kagome lattice are overlaid on top of highly resolved  Mn$_{3}$Ge (0001) surface. The Bright spots are assigned to Ge atomic sites and the dark spots refer to the triangular sub-lattice.
    (c) Atomically resolved image (\(7 \times 7 \, \text{nm}^2\)) taken in reversed biased(empty states) conditions ($V_{Bias}$ = 100 mV, $I_{t}$ = 850 pA). (d) A similar zoomed-in image of (c) (\(3 \times 3 \, \text{nm}^2\)), shows the reversal of LDOS from Ge sites to Mn spin triangles.
    (e) Line profile in (a) showing the  Ge-Ge periodicity on the a-b plane. The distance between two peak positions (0.52 nm) refers to the Ge-Ge interatomic distance.
    (f) The differential conductance(dI/dV) vs bias voltage measured on Mn and Ge atomic positions, along the red arrow indicated in (c), reveals a local dip near the Fermi level. The local minimum is likely to be originated from the Dirac crossings and multiple Weyl crossings near the Fermi level. (g) Tunneling spectra recorded at the Mn-rich area along the green line in c.}
    \label{fig:break3}
\end{figure*}

 For a better understanding of the change in surface morphology, we study the STM topographs for three batches of samples grown between $450^{\circ}$ C to $650^{\circ}$ C as shown in \hyperref[fig:break2]{\textcolor{blue}{Fig.~\ref*{fig:break2}}}.
At $450^{\circ}$ C, film morphology shows terraced granular-like structures within a scan area of 20 nm$^2$, as shown in \hyperref[fig:break2]{\textcolor{blue}{Fig.~\ref*{fig:break2}(a)}}. The lateral dimensions of the crystallites vary between 5-10 nm. The height variation of grains is measured along the line profile as shown in \hyperref[fig:break2]{\textcolor{blue}{Fig.~\ref*{fig:break2}(a)}}. The average terrace height ranges between 1-1.5 {\AA}. The maximum height of the crystalites is 2 {\AA} ,which is roughly the interplanar distance  between A-B stackings along the c-axis. 
In this stage the films surface features a higher number of  nucleation points leading to a set of independent crystallites with well-defined facets. The crystallites  exhibit three-dimensional growth due to the insufficient mobility of adatoms, which prevents them from diffusing beyond the grain boundaries before the next nucleation begins. The ability of atoms to overcome the Ehrlich–Schwoebel (ES) barrier is suppressed at lower growth temperatures, leading to the nucleation of adatoms on the existing islands \cite{kaufmann2016critical}. 
 
 The STM image in \hyperref[fig:break2]{\textcolor{blue}{Fig.~\ref*{fig:break2}(b)}} reveal a lower nucleation density at $550^{\circ}$ C. There is no significant enhancement in the terrace width but a near-ideal two-dimensional growth is established at this temperature range. Additionally, no secondary  nucleation sites are observed  over the existing terraces. The individual terraces develop a smooth morphology at the termination without entering into a multilevel growth regime. The islands develop a height of 2.21 \AA, as revealed from the height profiles in \hyperref[fig:break2]{\textcolor{blue}{Fig.~\ref*{fig:break2}(d)}}. The height difference correlates to the distance of two consecutive Kagome layers along the c-axis. This suggests that the films follow a layer-by-layer coverage, which can be attributed to the relatively high thermal mobility at $550^{\circ}$ C. Based on the morphological evidence from STM, it is expected to observe higher coverage of islands at higher growth temperatures in the unit cell regime. \hyperref[fig:break2]{\textcolor{blue}{Fig.~\ref*{fig:break2}(c)}} represent a large scale STM topography image on a film grown at $650^{\circ}$ C. The film shows distinct terraces with thicknesses of 2.15 \AA~\&~2.25 \AA~corresponding to A-B stacking, respectively \cite{yang2019scanning}. The average terrace width, as shown in \hyperref[fig:break2]{\textcolor{blue}{Fig.~\ref*{fig:break2}(c)}}, extends up to 100 nm at this regime. The improved layer-by-layer growth with increasing temperature is a direct consequence of enhanced diffusion length, which enables the adatoms to hop from a higher terrace to a lower terrace. The atomic resolution of the terrace reveals the hexagonal arrangement of the (0001) surface as shown in \hyperref[fig:break3]{\textcolor{blue}{Fig.~\ref*{fig:break3}(a)}}.  
The honeycomb lattice shows a  periodicity of 0.52 nm, which is consistent with Ge atomic sites in the (0001) plane. In order to reveal the Kagome feature of the (0001) surface, a bias-dependent study is conducted as shown in \hyperref[fig:break3]{\textcolor{blue}{Fig.~\ref*{fig:break3}(a)}} and \hyperref[fig:break3]{\textcolor{blue}{Fig.~\ref*{fig:break3}(c)}}. At a negative bias of -125 mV, the  brighter spots correspond to the Ge atomic sites of a Kagome plane. The darker spots  correspond 
to the triangular sublattice formed by the Mn atoms as shown in  \hyperref[fig:break3]{\textcolor{blue}{Fig.~\ref*{fig:break3}(b)}}. Evidently, at  -125 mV, the LDOS is intensified  at  the Ge sites. However, under a positive bias of +100 mV, LDOS is  concentrated at the center of Kagome triangles.

Furthermore, we measure the differential conductance (dI/dV) at different sites along the red and green lines in
\hyperref[fig:break3]{\textcolor{blue}{Fig.~\ref*{fig:break3}(f)}} and \hyperref[fig:break3]{\textcolor{blue}{Fig.~\ref*{fig:break3}(g)}} respectively. The green and red lines indicate the Mn-rich and undoped regions of the film respectively. Several representative tunneling spectra along the red line exhibit a dip in local density of states near the Fermi level. The v-shaped spectra is attributed to the linear gapless surface states observed in other Dirac materials \cite{jiang2012fermi,yang2019scanning}. However, Mn$_{3}$Ge develops  multiple Weyl points at the first Brillouin zone due to reduced SOC. The closely spaced Weyl points near the Fermi level convolute to one local minima and generate the v-shaped spectra. The certain drop in LDOS is attributed to the  band crossings near Weyl points. The spectra taken on Mn-rich sites in \hyperref[fig:break3]{\textcolor{blue}{Fig.~\ref*{fig:break3}(g)}} shares similar features with the undoped regions. The presence of local minima across all the atomic sites confirms the semi-metallic intersection of  conduction and valence bands near Fermi level. Substitution of nonmagnetic Ge atoms with Mn is expected to alter the band crossings near Fermi level. This also indicates that the topological properties of $Mn_{3}Ge$ are retained in thin films at this particular growth condition.
Various noncollinear antiferromagnetic arrangements of Mn$_3$Ge magnetic moments have been studied, as depicted in \hyperref[fig:break6]{\textcolor{blue}{Fig.~\ref*{fig:break6}(a-d)}}. Several DFT studies predict different results for the lowest-energy magnetic arrangement \cite{Zhang_2013, Yang_2017, PhysRevB.96.224415, 10.1126/sciadv.1501870}. Our results reveal the arrangement shown in \hyperref[fig:break6]{\textcolor{blue}{Fig.~\ref*{fig:break6}(a)}} to have the lowest energy, consistent with Refs.~\cite{Zhang_2013, Yang_2017}. The electronic band structure without and with spin-orbit interaction and the total density of states considering spin-orbit interaction associated with the energetically preferred antiferromagnetic configuration are displayed in \hyperref[fig:break6]{\textcolor{blue}{Fig.~\ref*{fig:break6}(e)}}, \hyperref[fig:break6]{\textcolor{blue}{Fig.~\ref*{fig:break6}(f)}}, and \hyperref[fig:break6]{\textcolor{blue}{Fig.~\ref*{fig:break6}(g)}}, respectively. Multiple band crossings are visible near the Fermi level, most of them having an apparent Weyl nature, while a Dirac crossing is seen at the high-symmetry K point, as circled in \hyperref[fig:break6]{\textcolor{blue}{Fig.~\ref*{fig:break6}(f)}}. The corresponding energy dispersion as a function of ($k_x, k_y$) is plotted and shown in \hyperref[fig:break6]{\textcolor{blue}{Fig.~\ref*{fig:break6}(h)}}, revealing two Dirac crossings at K and K$^{\prime}$ points, marked with circles.

\begin{figure*}
    \centering
    \includegraphics[width=1.0\textwidth]{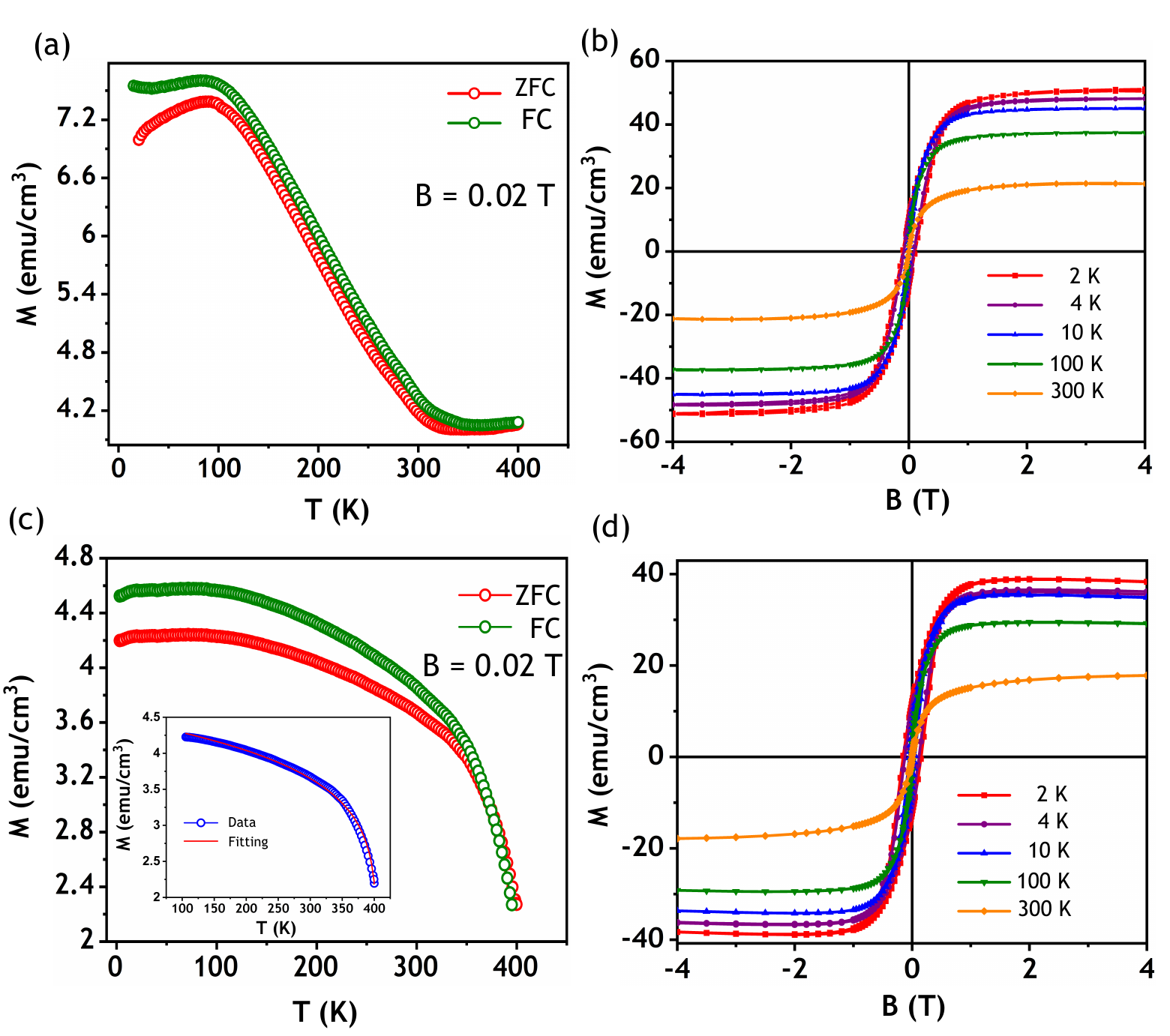} 
    \caption{ In-plane magnetization measurements on 30 nm thick $Mn_3Ge$ films grown at $550^{\circ}$ C and $650^{\circ}$ C. (a) Temperature-dependent magnetization M(T) measured in  zero field cooling (ZFC) and field cooling (FC) obtained at a field of 0.02 T. The ZFC-FC  curves shows a transition near 350 K, which is also the bifurcation point.
    (b) Field-dependent magnetization M(H) curves measured at five representative temperatures between 2 and 300 K. The uncompensated moment at 300 K is approximately 21 emu/$cm^3$. 
 (c) M(T) curves measured on films grown at $650^{\circ}$ C show similar splitting of the magnetization curves, but up to 400 K. The AFM Neel temperature ($T_{N}$ = 381 K) is estimated from the fitting of ZFC curve using \textcolor{blue}{eqn~(\ref{eq:eq1})}.(d) Isothermal M(H) loops measured on a similar set of temperatures between 2-300 K. A small uncompensated moment (18 emu/$cm^3$) at 300 K, is suggestive of a long range chiral order on films grown at $650^{\circ}$ C.}
\label{fig:break4}
\end{figure*}

\subsection{DC Magnetic Measurements}
 Large scale STM topographs reveal a layer by layer growth with long range Kagome order, also observed through bias dependent studies.  The establishment of long-range  NCAFM magnetic order in thin films is challenging and can be achieved only in a layer-by-layer mode of growth. To properly understand the magnetic properties in thin films, we performed magnetometry measurements on four sets of samples synthesized at different growth temperatures. Bulk Mn$_{3}$Ge shows a very high AFM Neel temperature $T_{N}$ = 390 ± 10 K with an intrinsic exchange bias persisting till room temperature \cite{qian2014exchange,yamada1988magnetic}. However, the magnetic properties of the topological Mn$_{3}$Ge thin films are not yet addressed completely \cite{wang2022investigating,wang2021robust}.  
 The temperature variation of magnetization (M-T) measured  in ZFC-FC mode is shown in \hyperref[fig:break4]{\textcolor{blue}{Fig.~\ref*{fig:break4}(a)}} and \hyperref[fig:break4]{\textcolor{blue}{Fig.~\ref*{fig:break4}(c)}}. 
In the ZFC mode, the sample is  initially cooled down to 2 K and data were collected while heating in the presence of the applied field. On the other hand, in the FC mode, samples are precooled in the presence of a field. 
 The ZFC-FC curves as shown in \hyperref[fig:break4]{\textcolor{blue}{Fig.~\ref*{fig:break4}(a)}}, are measured on a films deposited at $550^{\circ}$ C. The irreversibility in the ZFC and FC curves originates from  the low magnetic anisotropy of the Kagome lattice \cite{qian2014exchange,tomiyoshi1986triangular}. This enables the Mn spin triangles to rotate freely along the a-b plane, leading to a diversion in the ZFC-FC paths.  We observe an abrupt change in magnetization near 350 K. This transition is lower than the reported AFM Neel temperature ($T_{N}$). The magnetization curves of \hyperref[fig:break4]{\textcolor{blue}{Fig.~\ref*{fig:break4}(c)}} are acquired on samples grown at $650^{\circ}$ C. The bifurcation point of the ZFC-FC curve approaches 400 K, as observed in bulk single crystals. To precisely estimate $T_{N}$, the ZFC curve is fitted into the power law in \hyperref[fig:break4]{\textcolor{blue}{Fig.~\ref*{fig:break4}(c)}} inset.
\begin{equation}
 M(T) = M_{0}(1 - T/T_{N})^{\beta}
 \label{eq:eq1}
\end{equation}
where $M_{0}$ is the magnetic moment at $T = 0$ K. $\beta$ is an exponent, and  $T_{N}$ is the ordering temperature. The extracted value of  $T_{N}$ is  381.2 K, which is in close agreement to the previously reported values \cite{qian2014exchange}.

Discontinuous grain boundaries observed between $350^{\circ}$ C and $450^{\circ}$ C restrict the free rotation of magnetic spin triangles. This leads to a complete convergence of ZFC and FC paths, as demonstrated in the supplementary section (Fig. S2).  Substituting Mn atoms disrupts the chiral order in the Kagome lattice, reducing the magnetic ordering temperature (\(T_\mathrm{N}\)) of samples grown between \(350^{\circ}\,\text{C}\) and \(550^{\circ}\,\text{C}\) \cite{ghosh2023tuning}.
 Typical in-plane magnetic hysteresis of films grown at $550^{\circ}$ C and  $650^{\circ}$ C  are depicted in \hyperref[fig:break3]{\textcolor{blue}{Fig.~\ref*{fig:break4}(b)}} and \hyperref[fig:break4]{\textcolor{blue}{Fig.~\ref*{fig:break4}(d)}} respectively. The M-H loops are obtained after a linear subtraction of the diamagnetic background of the substrate.
 
\begin{figure*}
    \centering
    \includegraphics[width=1.0\textwidth]{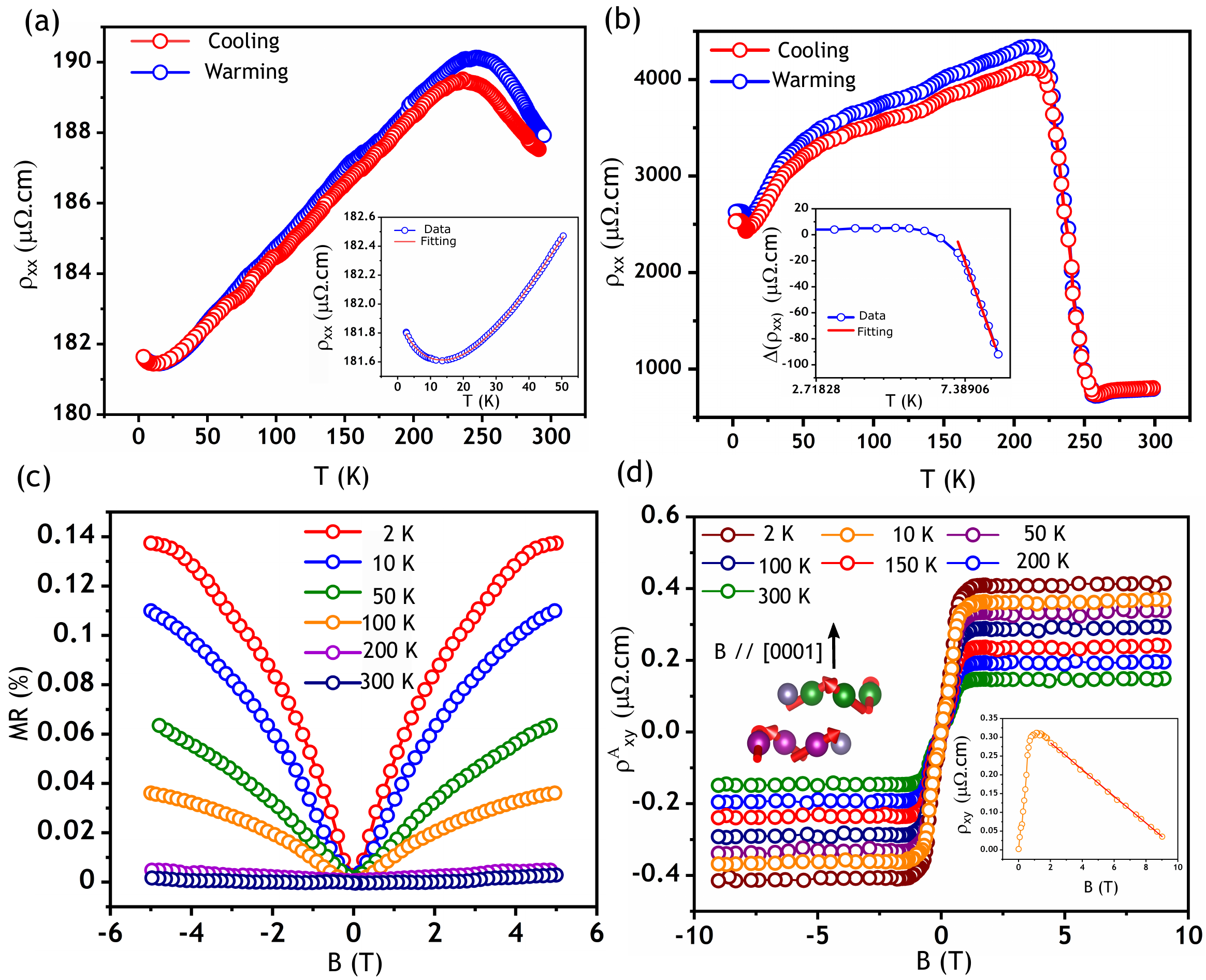} 
    \caption{ Magneto-transport studies on 30 nm thick $Mn_{3}Ge$ thin films grown at $650^{\circ}$ C and $550^{\circ}$ C. (a) and (b) Temperature variation of longituidinal resistivity ($\rho_{xx}$-T) measured on films grown at $650^{\circ}$ C and $550^{\circ}$ C respectively. The inset in (a) shows an enlarged view of the fitting between  2–50 K  using \textcolor{blue}{eqn~(\ref{eq:eq2})}. The inset in (b) is the fitting of low-temperature upturn using a semi-logarithmic plot. The logarithmic (-lnT) increase with decreasing temperature reflect the presence of Kondo effect.(c) Temperature dependent transverse magnetoresistance (MR) measurement within a field range of $\pm$5 T. Linear MR response in the lower fields can be attributed to the linear dispersion of electronic states. (d) Variation of anomalous Hall resistance ($\rho^{A}_{xy}$)  with magnetic field (B) at  seven representative temperatures between 2-300 K. Inset(d) shows the extraction of $\rho^{A}_{xy}$  from $\rho_{xy}$  by subtracting the linear component ($\rho_H B$) in the high field regime. Schematic illustrational of the Mn magnetic moment and formation of spin Berry phase.  }
    \label{fig:break5}
\end{figure*}
The magnetization increases with field initially in the low-field regime, till it reaches a saturated magnetized state at $H_{s}\approx$ 1.5 T. The isotherms exhibit negligible spontaneous magnetization at the zero field regime at all the measured temperatures. The saturated magnetization ($M_{s}$) decreases with increasing temperature, indicating a transition from a chiral magnetic order to a weak ferromagnetic state. The Mn-rich films grown at $550^{\circ}$ C show a slightly higher value of $M_{s}$ = 21 $emu/cm^3$ at 300 K, as compared to 18.2 $emucm^3$ for $650^{\circ}$ C \cite{zhao2021magnetic,wang2022noncollinear}. The ferromagnetic behaviour originates from the uncompensated moments of the triangular sublattice of the a-b plane. The magnetic moment per Mn atom is estimated to be 0.047 $\mu_\text{B}$(0.03 $\mu_\text{B}$) at $550^{\circ}$ C($650^{\circ}$ C) respectively, which is almost an order higher than its bulk state (0.005 $\mu_\text{B}$)\cite{ogasawara2019structural,nayak2016large}. The weak ferromagnetic behavior and the simultaneous presence of Weyl crossings near the Fermi level confirm that films grown at $650^{\circ}$  are in a magnetic-Weyl semimetallic state.

\subsection{Magneto-transport measurements}
To  probe the electrical signatures of magnetic Weyl semimetal, we measure the temperature dependence of electrical resistivity ($\rho_{xx}$ -T) on the same sample. Referring to \hyperref[fig:break5]{\textcolor{blue}{Fig. \ref*{fig:break5}(a)}}, $\rho_{xx}$ exhibits metallic behavior within the temperature range of 2 to 220 K. Beyond 220 K, it follows a slightly downward trend attributed to the metal-semimetal transition \cite{rai2022unconventional}. 
\begin{figure*}
    \centering
    \includegraphics[width=1.0\textwidth]{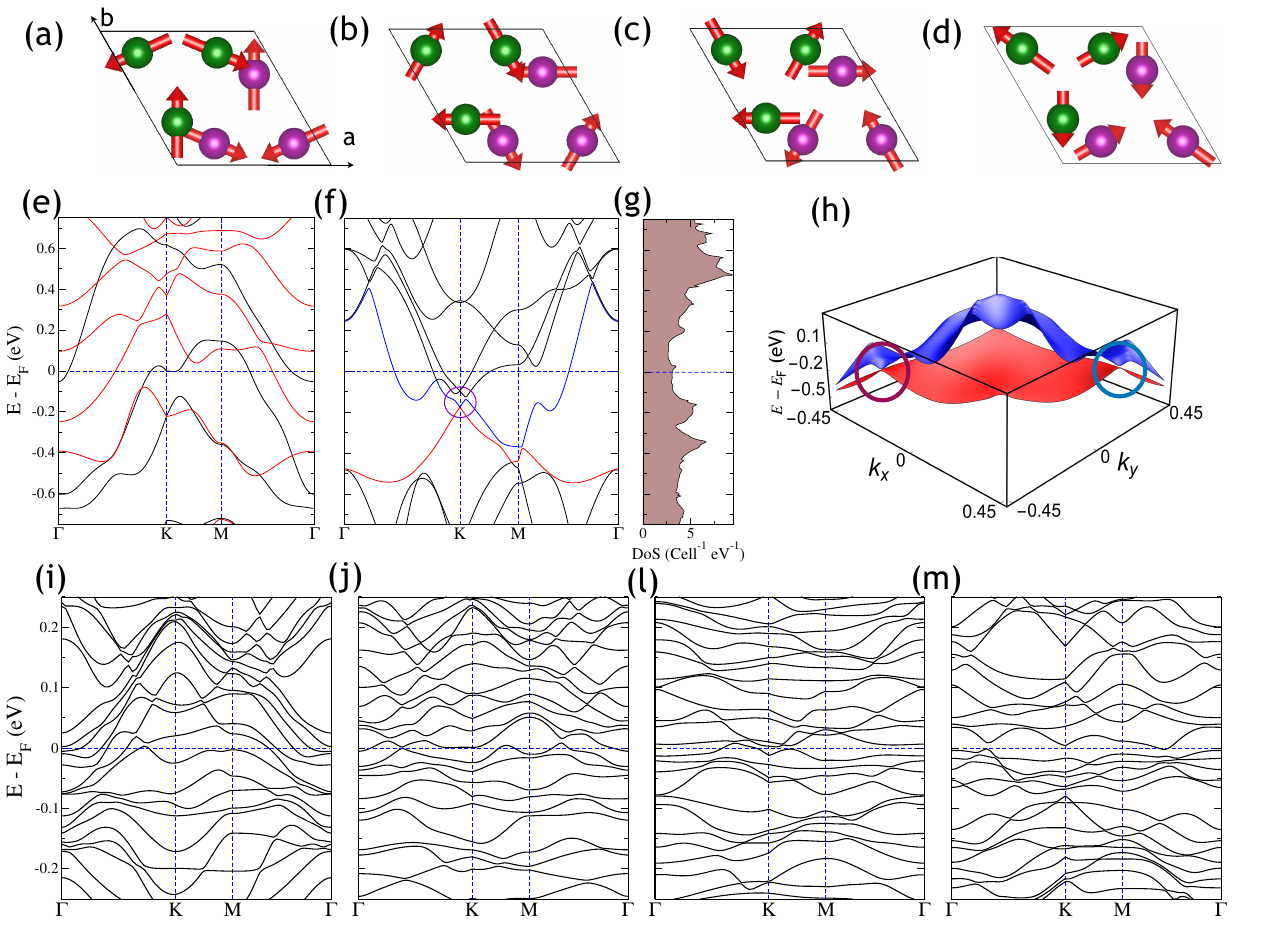} 
    \caption{A few possible chiral antiferromagnetic arrangements of Mn$_3$Ge and the electronic structure of the lowest-energy arrangement are displayed here. Panels (a), (b) and (c), (d) show antiferromagnetic arrangements with opposite spin chirality. The structure images are prepared using the {\scshape vesta} software \cite{momma2011vesta}. The magenta and green atoms correspond to Mn at Wyckoff position 6h with z = 1/4 and 3/4, respectively. Ge atoms are omitted for clarity. Spin polarized bandstructure for the lowest energy magnetic arrangement without and with spin-orbit interaction are shown in (e) and (f), respectively, for a conventional unit cell. Panel (g) shows the total density of states of Mn$_3$Ge within spin-orbit interaction. A part of the valence and conduction bands are visualized in 3D by plotting the energy as functions of $k_x$ and $k_y$, keeping $k_z = 0$, as displayed in (h), where the Dirac cones at K and K$^{\prime}$ points are circled. The band dispersion of Mn$_{3+x}$Ge$_{1-x}$ with increasing doping concentration $x$ including spin-orbit interaction are shown in panels (i), (j), (l), and (m) for 5.55\%, 11.11\%, 27.78\%, and 33.33\% doping concentrations, respectively.}
    \label{fig:break6}
\end{figure*}
 To understand the scattering mechanism at the low temperature, $\rho_{xx}$ have been fitted to the following empirical equation:
\begin{equation}
\rho_{xx} = \rho_{0} + \alpha T^2 + \beta T^3 + \gamma T^{0.5}
\label{eq:eq2}
\end{equation}
The coefficients $\alpha$, $\beta$ and $\gamma$ correspond to (e-e) scattering, magnon scattering, and weak localization \cite{wu2022structural}. The values of $\rho_0 = 181.7 \, \mu\Omega \, \text{cm}$, $\alpha = 9.3 \times 10^{-4} \, \mu\Omega \, \text{cm}/\text{K}$, $\beta = 6.4 \times 10^{-6} \, \mu\Omega \, \text{cm}/\text{K}^2$, and $\gamma = -0.1659 \, \mu\Omega \, \text{cm}$ are obtained from the fitted model as shown in \hyperref[fig:break5]{\textcolor{blue}{Fig.~\ref*{fig:break5}(a)}}. According to the relative weight of fitting parameters, Weak localization and e-e scattering dominate the scaterring at low temperature. 
Films grown at  $550^{\circ}$C exhibit a broader transition near 220 K as shown in \hyperref[fig:break5]{\textcolor{blue}{Fig.~\ref*{fig:break5}(b)}}. The low-temperature upturn is attributed to the Kondo effect arising from excess Mn doping. The logarithmic increase in electrical resistivity in \hyperref[fig:break5]{\textcolor{blue}{Fig.~\ref*{fig:break5}(b)}} inset, is a characteristic signature of the Kondo effect \cite{qin2020anomalous}. Substituted Mn spins exchange-coupled to the 3d bands induce Kondo scattering through interaction with conduction electrons. Kondo effect has been theoretically predicted in magnetically doped 3D Dirac and Weyl semimetals \cite{mitchell2015kondo,yanagisawa2015kondo}. Recent experimental studies also confirm the coexistence of Kondo effect and Weyl  states in Mn-doped compounds \cite{lee2022coexistence,khadka2020kondo,kurosawa2022chiral}. 

\hyperref[fig:break5]{\textcolor{blue}{Fig.~\ref*{fig:break5}(c)}} shows the isothermal MR studies at different temperatures with a magnetic field parallel to the c-axis(0001).
The Hall Contribution was removed by averaging the $R_{xx}$ data for positive and negative magnetic fields. The change in magnetoresistance (MR) is calculated using the formula $MR = \left[\frac{R(B) - R(0)}{R(0)}\right] \times 100\%$, where  R(B) is the value of longitudinal resistance at the field (B). 
As evident from the MR plot a linear dependence of MR is observed in the low-field region. There are primarily two mechanisms behind the linear dominating term \cite{low2024anisotropic}. (1)The existence of linear dispersive Dirac like bands that have a remarkably low effective mass close to the Fermi level \cite{abrikosov1998quantum,feng2015large}.(2) Fluctuations in carrier density due to inhomogeneities or grain boundaries non-disordered systems. As we have already characterized our films to be epitaxial and continuous, we have also eliminated mechanism 2. Films grown at $550^{\circ}$C exhibit negative magnetoresistance as shown in supplementary section (Fig. S3). Negative MR may result from reduced spin-dependent scattering, as the magnetic field aligns the Mn spins at the hexagonal center. Next, we measure isothermal Hall resistivity at different temperatures as shown in \hyperref[fig:break5]{\textcolor{blue}{Fig.~\ref*{fig:break5}(d)}} . The magnetic field is applied along the (0001) direction, perpendicular to the Kagome plane.
The contacts were made in Hall geometry and are often accompanied by probe misalignment. To reduce such additional contributions to Hall resistivity, $\rho_{xy}$ is evaluated with its antisymmetric term using the relation $\rho_{xy}(B) = \frac{1}{2} \left(\rho_{xy}(+B) - \rho_{xy}(-B)\right)$.  
 The standard expression for the transverse resistivity $\rho_{xy}(B)$ for typical ferromagnets is given by 
\begin{equation}
    \rho_{xy}(B) = \rho_H B + 4\pi \rho_s M
    \label{eq:eq3}
\end{equation}

 where $\rho_{H}B$ represents the ordinary Hall effect (OHE) caused by the Lorentz effect and the second term $4\pi \rho_s M$ represents the contribution of the anomalous Hall effect ($\rho_{xy}^{AHE}$). As the magnetic field increases, $\rho_{xy}$ exhibits a sudden change in slope, indicating the presence of AHE.   
 The Hall slope ($\rho_H$) is estimated by fitting the high field region of the ($\rho_{xy}$) graph as depicted in the inset of \hyperref[fig:break5]{\textcolor{blue}{Fig.~\ref*{fig:break5}(d)}}. The estimated carrier concentration using the hall slope at 10 K is around $(\sim 10^{23} \, \text{cm}^{-3})$  justifying the metallic behavior of Mn$_{3}$Ge films. The anomalous Hall term $4\pi \rho_s M$ is determined by subtracting the linear component of the $\rho_{xy}$ in the high magnetic-field regime. \hyperref[fig:break5]{\textcolor{blue}{Fig.~\ref*{fig:break5}(d)}}  shows the variation of $\rho_{xy}^{AHE}$ with magnetic field at different temperatures. $\rho_{xy}^{AHE}$ and $\sigma_{xy}^{AHE}$ are usually suppressed  due to lower spin Berry phase under B $\parallel$[0001] configuration \cite{kiyohara2016giant,nayak2016large}. The observed values of $\rho_{xy}^{AHE}$ and $\sigma_{xy}^{AHE}$ at 2 K are around 0.41 ($\mu\Omega\cdot\text{cm}$) and 11.5 ($\Omega^{-1}\cdot\text{cm}^{-1}$) respectively. The obtained values of $\rho_{xy}^{AHE}$ and $\sigma_{xy}^{AHE}$ are comparable to the sputtered grown films and close to bulk crystals \cite{nayak2016large,takeuchi2024magnetic}. $\sigma_{xy}^{AHE}$ is calculated from ( $\sigma_{xy}^{AHE} = \frac{-\rho_{xy}^{AHE}}{\rho_{xx}^2}$
) and the temperature dependence is shown in the supplementary section (Fig. S4).    
 The observed  $\rho_{xy}^{AHE}$ in B $\parallel$[0001] configuration is likely attributable to real-space spin chirality as shown in \hyperref[fig:break5]{\textcolor{blue}{Fig.~\ref*{fig:break5}(d)}}. The triangular spin configuration, when subjected to a [0001]-oriented magnetic field develops a spin canting with nonzero scalar spin chirality $(\mathbf{S}_1 \cdot (\mathbf{S}_2 \times \mathbf{S}_3)$ \cite{khadka2020kondo,ghosh2023tuning}. Under an external magnetic field, the Dzyaloshinskii-Moriya interaction (DMI) induces an imbalance in spin populations with opposite chiral domains, resulting in a nonzero Berry curvature. When the magnetic field is reversed, the chirality of the spin population also switches, leading to a reversal in the Hall resistance \cite{rai2022unconventional}. To confirm that AHE is originating from the spin chirality, we have performed comparative (M-H) loop measurements for in-plane and out-plane geometries. The saturation moment measured in the out-of-plane configuration is an order smaller than the in-plane moments, as shown in the supplementary section (Fig. S5). This also indicates the small tilting of Mn spins along the c-axis. This suggests that $\rho_{xy}^{AHE}$ is insensitive to magnetization and is related to the chiral spin structure, as predicted by theory.  
 In the case of lower growth temperatures ($550^{\circ}$ C), $\rho_{xy}-B$ shows a linear response with a magnetic field as shown in the supplementary section (Fig. S3(c)). However, this could be driven by certain factors like stochiometry as Mn atoms occupying Ge positions can suppress the overall Berry curvature by annihilating the Weyl nodes. The vanishing AHE could also originate from the grain boundary discontinuities that hinder the long-range interaction of triangular Mn moments.
 \subsection{DFT calculations}
 A strongcorrelationn between band structure and film composition is provided by the magnetotransport studies, which show Mn substitution increases the resistivity, driving $Mn_{3}Ge$ into a semiconducting and non-topological state. As $\rho_{xy}^{AHE}$  also depends on the band structure, we carried out DFT calculation of the electronic structure of Mn$_{3+x}$Ge$_{1-x}$ with spin-orbit interaction induces a substantial and collective change in electronic, magnetic and topological properties. We simulate Mn-doping in the system by substituting Ge atoms, forming Mn$_{3 + x}$Ge$_{1 - x}$, with increasing doping concentration $x$. The band dispersion with increasing doping concentration results in attenuation of topological features with diminishing states near the Fermi level.
 \hyperref[fig:break6]{\textcolor{blue}{Fig.~\ref{fig:break6}(i-m)}}
 presents the band structure for Mn$_{3+x}$Ge$_{1-x}$ at $x =$ 5.55\%, 11.11\%, 27.78\%, and 33.33\% along the $\Gamma -K-M- \Gamma$ direction. The conduction band remains mostly unaffected with low concentration of doping, while moderate to high doping concentration a gap is introduced at the Fermi level as shown in \hyperref[fig:break6]{\textcolor{blue}{Fig.~\ref*{fig:break6}(i-m)}}. The energy splitting between the valence band and conduction band increases with increasing Mn concentration, leading to the emergence of a small band gap at the Fermi level with significant concentrations of Mn. This results in a substantial reduction in the bandwidth near the Fermi level.

\section{Conclusion}
Highly c-axis oriented $Mn_{3}Ge$ thin films with the pure hexagonal structure were successfully synthesized on Si(100) substrate using pulse laser deposition. Our detailed X-ray scattering and atomic force microscopy studies establish a suitable growth condition to obtain layer-by-layer films with flat surfaces and well-defined interfaces. High-resolution STM study on $Mn_{3}Ge$ thin films grown at different temperatures, revealed the detailed mode of growth and surface surface atomic structures, where the Mn atoms spontaneously arrange into a Kagome lattice. Tunneling spectra measured on the surface indicate local minima originating from the Weyl crossings near the Fermi level, consistent with the first principle electronic structure calculations. Temperature-dependent magnetization and transport measurements showed Berry curvature-induced large AHE that persists up to room temperature. Finally, Our first-principles DFT calculations indicate that the stoichiometric films possess strongly correlated metals and with increasing concentrations of Mn on Ge sites induce band gap near the Fermi level. 

\section{Data availability statement}
The authors declare that the data that support the findings are available within the article and supplementary section.

\section{Acknowledgment}
 This work was supported by the Anusandhan National Research Foundation (ANRF) under SERB Core Research Grant(CRG/2023/008193). I.S is grateful for fellowship assistance from the MHRD, Government of India. All authors acknowledge the Department of Physics, IIT Delhi for providing XRD, PLD, MPMS and PPMS facility. All authors sincerely thank Amod Kumar, Subham Naskar and Saurav Sachin for their technical help. P.D.\ acknowledges UGC, India, for the research fellowship [grant no. 1498/(CSIR NET JUNE 2019)] and the high-performance computing facility at IISER Bhopal. N.G.\ acknowledges SERB, India for research funding through grant number CRG/2021/005320.
 
\bibliography{Isinha_Main.bib}

\end{document}